# DEEP LEARNING MEETS BLOCKCHAIN FOR AUTOMATED AND SECURE ACCESS CONTROL

Asma Jodeiri Akbarfam[1], Sina Barazandeh[2], Deepti Gupta[3], and Hoda Maleki[1]

[1] School of Computer and Cyber Sciences, Augusta University, Augusta, GA, USA
[2] Department of Computer Engineering, Bilkent University, Ankara, Turkey
[3] Subhani Department of Computer Information Systems, Texas A&M University-Central Texas, TX, USA

*ABSTRACT*

*Access control is a critical component of computer security, governing access to system resources. However, designing policies and roles in traditional access control can be challenging and difficult to maintain in dynamic and complex systems, which is particularly problematic for organizations with numerous resources. Furthermore, tra- ditional methods suffer from issues such as third-party involvement, inefficiency, and privacy gaps, making transparent and dynamic access control an ongoing research problem. Moreover detecting malicious activities and identifying users who are not behaving appropriately can present notable difficulties. To address these challenges, we propose DLACB, a Deep Learning Based Access Control Using Blockchain, as a solution to decentralized access control. DLACB uses blockchain to provide transparency, traceability, and reliability in various domains such as medicine, finance, and government while taking advantage of deep learning to not rely on predefined policies and eventually automate access control. With the integration of blockchain and deep learning for access control, DLACB can provide a general frame-work applicable to various domains, enabling transparent and reliable logging of all transactions. As all data is recorded on the blockchain, we have the capability to identify malicious activities. To expedite this process, we store a list of ma- licious activities in the storage system and employ a verification algorithm to cross-reference it with the blockchain. We conduct measurements and comparisons of the smart contract processing time for the deployed access control system in contrast to traditional access control methods, determining the time overhead involved. The processing time of DLBAC demonstrates remarkable stability when exposed to increased request volumes. This approach is particularly useful for organizations seeking to automate access control while simultaneously detecting and preventing data breaches to enhance security.*

## 1. INTRODUCTION

Blockchain is a peer-to-peer network [40] and a method of storing information. It prevents unwanted modification and manipulation of data while providing a decentralized ledger for transactions. It is also used for authenticating, maintaining, and synchronizing the content of a transaction ledger replicated across multiple users. Blockchain facilitates decentralized transactions and data management and offers an environment where no third party controls the data, and no trust is required between stakeholders [2]. The features of blockchain make it tamper-resistant, providing integrity and a transparent view of what is happening to the data [13,59]. Therefore, blockchain is useful in various domains for data management such as Internet of Vehicles (IoV), energy systems, and supply chain management [55] or for ensuring security and privacy [2,39] such as medical systems [23], internet of things [31], finance [21] and access control [51].

In the context of access control, blockchain technology can be used to record and store access control policies and access requests [18]. Each subject can have their own blockchain-based identity that contains their attributes and access rights. When an access request is made, the policy is evaluated against the access context, and the decision is recorded on the blockchain. This ensures transparency and audibility of access decisions. Blockchain technology enables secure and decentralized transfer of access rights. Instead of relying on a central authority to transfer access rights, blockchain allows for the peer-to-peer transfer of access rights through smart contracts. Decentralization of access control eliminates the need for intermediaries and reduces the risk of fraud [20].

Traditional approaches to access control, such as Access Control Lists (ACLs) [10], Role- Based Access Control (RBAC) [16] , and Attribute Based Access Control (ABAC) [22] have been utilized in blockchains [18,36,44]. However, these types of access control have limitations such as over-provisioning and under-provisioning of access, complexity in policy and attribute engineer- ing, poor generalization to new situations, and utilizing a large amount of memory in the system [41]. Therefore, there is an increasing demand for innovative access control approaches, including the utilization of deep learning-based models. Although few studies [33,42,15] have proposed deep learning-based access control methods, they have not yet been integrated with blockchain technology. These approaches can automate and dynamically manage access control decisions. However, they may face challenges such as a central point of failure or lack of transparency, making them unsuitable for organizations to detect and prevent data breaches and insider attacks. To overcome these challenges, there is a need for new approaches that provide a secure and transparent access control mechanism.

We propose Deep Learning Based Access Control Using Blockchain (DLACB), which com-bines the benefits of blockchain and deep learning to provide a secure and transparent access control mechanism. DLACB relies on smart contracts to verify users and integrates a deep learn- ing model for precisely assessing each user's access level within the system. The utilization of deep learning automates access control and eliminates the need for policy development. Addition- ally, the framework employs smart contracts to consider organizational priority rules, detect ma- licious users, and enforce bans when necessary. The framework is based on a private blockchain, which only allows trusted nodes to validate transactions. This ensures the security and integrity of the access control mechanism by preventing unauthorized access to the blockchain while not re- quiring third-party authentication service. In addition, Each user has their own blockchain-based identity, which ensures that only authorized users can access the system resources. DLACB al- lows users to make access requests by generating new transactions, which are then logged on the blockchain. This enables access transparency and helps detect malicious user activities in retriev- ing resources. Moreover, we have incorporated a simple method to lock out malicious users after their initial attempt and maintain an easily retrievable log of malicious activities stored securely in the system. To further ensure the accuracy of this log, we employ a verification method that leverages blockchain technology, offering fast retrieval of malicious activities while preserving data integrity.

The integration of blockchain and deep learning in DLACB creates a resilient and secure access control framework. Utilizing blockchain technology ensures transparency, authentication, immutability, detection of malicious users, and secure communication with users and storage. Simultaneously, the incorporation of deep learning enhances access control by enabling automated decision-making, adapting to dynamic environments, improving accuracy, generalizing to new scenarios, and minimizing human error.

The main contributions of this paper are as follows.
– We identify a research gap in traditional access control models, which rely on predefined

policies and permissions.

– We propose DLACB, which is a framework that utilizes deep learning access control to determine a user's permissions on a given resource while having the capability of adding organizational rules for access. The framework also authenticates the users and logs the access requests on the blockchain to recognize malicious users.

– We integrate a methodology for storing malicious user activities, retrieving them, and cross-referencing them with the blockchain for verification purposes.

– We present the blockchain communication process, such as consensus protocols and smart contracts to develop this framework and its cryptographic techniques, including hash func- tions, digital signatures, and encryption algorithms.

– We demonstrate the security analysis a perform measurements and comparisons of the smart contract processing times between the deployed access control system and traditional access control methods, aiming to ascertain the extent of time overhead incurred.

Our proposed framework can serve as a basis for the development of secure and efficient access control systems in various domains, including finance, healthcare, and government. By leveraging the power of Deep Learning and Blockchain technologies, our framework provides a robust and transparent solution to the critical issue of access control. The rest of this paper is structured as follows. Section 2 discusses an overview of blockchain and access control. Section 3 explores the related work. Section 4 discusses the importance of the method, describes the framework and the communication between entities. Section 5 describes the details of the protocols, Section 6 analyzes the security of the platform, Section 7 illustrates how the system was implemented and the details of the used deep learning model. Section 8 contains the evaluation. Section 9 delves into the proposed approach, providing insights into the reasons behind certain decisions made and future work. The paper concludes with section 10.

## 2. BACKGROUND

### 2.1. Blockchain

Blockchains are peer-to-peer networks and timestamped chains of blocks [3] as shown in Figure 1 maintained by participating nodes and transactions that are accumulated in blocks. Every block is chained to the previous block and cryptographically linked by including the previous block's hash value [12,5]. Blockchain provides immutable data storage as new blocks which can only be appended to the end of the chain, meaning existing transactions cannot be updated or deleted. Thus, transactions can be trusted without the assistance of any third parties. The immutable chain of historical transactions guarantees non-repudiation of historical transactions. In order to prove identity and authenticity and secure read and write access to the blockchain, cryptography primitives such as digital signatures, hash functions, and ciphers are used.

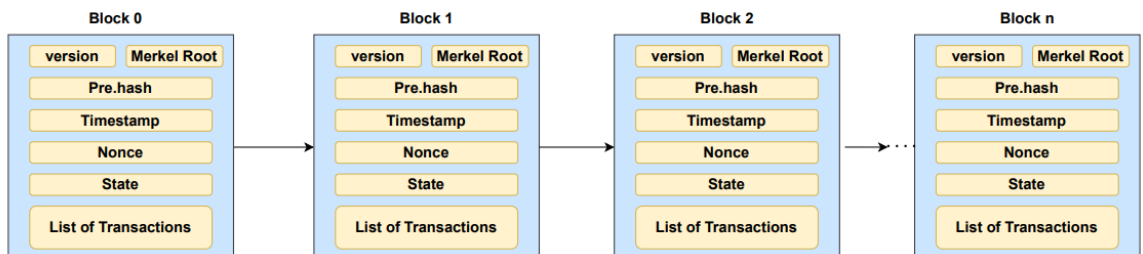

Fig. 1: Chain of Blocks

A vital aspect of a blockchain network is the consensus algorithm that ensures integrity and security. The consensus is a procedure that enables all participants in a distributed computing environment to reach an agreement on the ledger's current data state and be able to trust peers [54]. Furthermore, fraud and data tampering is prevented because data cannot be changed without the permission of a majority of the stakeholders in a blockchain system. Blockchain ledgers allow for sharing but not for modification.

There are various types of blockchains, including public and private [17,53,58]. A distributed ledger system without constraints and permissions is known as a public blockchain, with examples of Bitcoin and Ethereum. Anyone with internet access can join the network as an authorized node and become a part of the blockchain. It is permitted for a node or user who is a part of the public blockchain to view recent and old records, confirm transactions and engage in mining. The mining and trading of cryptocurrencies is the most fundamental usage of public blockchains.

A restricted or private blockchain [58,9] that may only be used in a closed network is a private blockchain. A private blockchain is typically utilized within businesses or organizations where only a small group of people are allowed to participate in the blockchain network. The governing organization controls the level of security, authorizations, permissions, and accessibility. In other words, private blockchains are similar to public blockchains, but with restricted access [32]. Private blockchains are mostly used for cases where centralized control of the network is needed, such as asset ownership, digital identity, supply chain management, voting, etc.

Some blockchains, including Ethereum-based ones, consist of smart contracts, which are modular, reusable, and automatically executed programs [38]. A smart contract responds to information sent to it, receives and stores the values, and transfers the information and values to other participants in the blockchain network. Once a smart contract is launched, the compiler generates a byte code that is stored on the blockchain, and it is assigned a storage address. All network nodes execute the opcodes produced by compiling the contract scripts when a condition specified in the contract takes place. The execution results are then written to the blockchain by the network nodes once a transaction is sent to the contract address.

## 2.2. Access Control

Access control is a crucial component of security that restricts access to resources on a system to authorized users while preventing unauthorized access. By implementing restrictions and permissions, access control reduces the risks of data breaches and other security threats to organizations or businesses. Security compliance programs rely heavily on access control policies and technology to protect sensitive information, such as customer data.

There are various traditional access control types, including attribute-based access control (ABAC), role-based access control (RBAC), and rule-based access control. In ABAC, every user has an attribute defined by the administrator, and their request can be accepted or denied based on these attributes [26,14,27]. RBAC defines the level of access based on the user's responsibilities and departmental jobs [49]. Rule-based access control restricts access to different parts of a system based on predefined rules for each area [24]. However, traditional access control techniques have some drawbacks, including the provision of excessive or insufficient access, challenges in creating policies and attributes, limited adaptability to new scenarios, and significant memory usage in the system [42].Consequently, there is an increasing demand for innovative access control methods, such as models based on deep learning, to overcome these limitations.

## 3. RELATED WORK

Blockchain technology has demonstrated its effectiveness in diverse sectors. In supply chain management, it enhances traceability, transparency, product authenticity, and combats counterfeiting, facilitating decentralized information sharing [4]. In healthcare, ensuring the security and privacy of electronic medical records remains a challenge [29]. blockchain ensures secure and efficient medical information exchange, preserving data integrity and confidentiality while addressing data management challenges [52]. For voting systems, blockchain guarantees secure, transparent, and immutable transactions, boosting the integrity of the process [28]. Blockchain also has the potential to be utilized with insurance solutions against political risks [57]. In recent years, there has been a significant increase in the adoption of blockchain technology in access control systems. This adoption provides a decentralized and secure framework for implementing access control policies. Within this section, we delve into various research endeavors that have put forward blockchain-based access control solutions tailored to different purposes and with the utilization of various types of access control.

*RBAC on blockchain:* In the context RBAC, several studies have explored its application within blockchain technology[16,47]. Sun et al. [50] introduced an RBAC model using a private blockchain. They utilized a multi-index table and a key-value database to formulate access control policies and store summarized access control data on the blockchain. Additionally, they employed a smart contract to handle user access queries and adopted the distributed proof of stake (DPOS) as the consensus algorithm within the blockchain. This private blockchain was combined with a Mandatory Access Control (MAC) policy. Kim et al. [30] utilized RBAC on a blockchain for shar- ing filmed images in closed-circuit television systems for crime prevention. Similarly, Cruz et al. The authors in [8] introduce an approach to secure access control in a private blockchain. It empha- sizes the initial deployment of a RBAC, to effectively categorize users based on predefined roles, restricting their access to specific parts of the blockchain. This is further fortified by the integration of a Machine Learning (ML) model, enabling the system to detect intricate user behavior patterns and potential security threats in real-time. [16] applied RBAC and defined a smart contract-based authentication method suitable for trans-organizational utilization of roles. However, traditional access control models like RBAC can lead to errors in policies and require significant memory for storing access lists, limiting their ability to handle unforeseen access requests. RBAC-based blockchain solutions may encounter problems related to policy engineering, memory usage, and difficulties in handling novel access requests.

*ABAC on blockchain:* In the realm of ABAC, blockchain has also found applications [48]. Given the swift evolution of the Internet of Things (IoT) [6], blockchain proves especially bene-ficial in this sphere [56], along with data-sharing situations. Ding et al. [19] and Zaidi et al. [56] employed blockchain for IoT access management using ABAC. Ding's approach stored attribute distribution on the blockchain, while Zaidi utilized smart contracts for fine-grained access control in IoT data sharing. This ABAC-based approach enables more flexible and granular access control. Abdi et al. [1] presented a hierarchical blockchain architecture for ABAC in the IoT environment, aiming to reduce network overhead and transaction latency. They introduced various types of managers and employed practical Byzantine fault tolerance (PBFT) as the consensus algorithm. This approach, while more robust, may require extensive attribute engineering. ABAC-focused blockchain solutions may face challenges related to attribute engineering and policy definition, potentially leading to time-consuming processes and policy errors.

*Reinforcement Learning and Dynamic Access Control on blockchain:* Outchakoucht et al. [43] introduced a dynamic and decentralized security framework that leverages blockchain tech- nology and reinforcement learning (RL) algorithms. This approach eliminates the need for cen- tralized authorities in access control and continuously adapts and optimizes security measures.

However, it exhibits limitations concerning blockchain privacy and block validation duration.

***Privacy-Preserving Healthcare Data Sharing:*** Blockchain has also been applied to secure healthcare data sharing. Madine et al. [34] proposed a decentralized data management model in which patients control their data, including the ability to remake re-encryption keys. Similarly, Alhajri et al. [7] introduced a human-centric decentralized consent system for fitness data sharing via blockchain. Zou et al. [59] presented a medical data sharing and privacy-preserving eHealth system (SPChain) using proxy re-encryption schemes to allow patients to share data while maintaining privacy. Challenges in healthcare data sharing solutions include key management, deployability, and authentication rules, as well as the risk of incorrect access decisions.

The DLACB framework stands out for its use of deep learning to enhance access control decision-making. It automates permission grants and eliminates the need for predefined policies, and policy engineering. Additionally, It helped with fast detection of malicious activity and adaptability to limiting malicious users. Overall, the research landscape in blockchain-based access control systems offers a spectrum of approaches, each with its own unique set of advantages and challenges.

## 4. DLACB FRAMEWORK

In organizations, outsourcing access control functions to third parties has become popular. This eliminates the need to configure and maintain complex systems that could incur high acquisition and operating costs [35]. These systems require log production that usually reveals the system's regular events and assists in detecting malicious behavior or attacks. Analyzing these logs helps to catch security breaches or incident discovery and subsequent damage control [46]. Therefore, these logs need to be immutable to prevent attackers from altering them and causing analysis to be incorrect.

This paper proposes the DLACB; a novel approach for implementing access control services by utilizing blockchain technology. Our framework creates immutable system logs and eliminates the need for trusting a third party while carrying the mentioned advantages. As part of our approach, an access control service is built on top of a private blockchain. It leverages blockchain technology for configuring access control policies based on a decision engine. In other words, the decision engine evaluates an access request for a resource This decision engine combines a deep learning model with priority rules. It utilizes deep learning for access control policies, facilitating informed decisions based on user and resource metadata across various operations. This approach enables dynamic adaptability, improved accuracy, reduced human errors, and enhanced overall efficiency, allowing it to seamlessly navigate evolving environments and effectively generalize to new scenarios. By leveraging deep learning for access determination, the reliance on predefined, rigid policies and rules is eliminated, ultimately enhancing the precision of access decisions. Additionally, although not mandatory, specific priority rules can be applied within the decision engine, tailored to conditions within a given network or organization, ensuring the effective adaptation of the access control policy to predefined circumstances.

### 4.1. Motivations and Objectives

DLACB offers several notable advantages over prior approaches:

1. **Elimination of the need for third-party trust:** Blockchain integration removes the necessity for third-party reliance, allowing for decentralized and transparent access control decisions verified through network consensus. The tamper-resistant and immutable ledger characteristic of blockchain enables autonomous user authentication, access rights management, and a secure environment without external dependencies.

2. **Enhanced Security:** DLACB's utilization of blockchain technology bolsters security, providing robust protection for sensitive data such as access control policies and user permissions.

3. **Streamlined Administrative Processes:** DLACB simplifies access control procedures, liberating administrators to allocate their resources to other areas.

4. **Adaptability:** DLACB's dynamic access determination powered by deep learning enables swift adaptation to new scenarios and emerging threats.

5. **Prevention of Replay Attacks:** DLACB incorporates preventive measures, including nonces, to thwart replay attacks. This ensures that each request or transaction is processed only once, reinforcing system security.

6. **Malicious Activity Detection:** The framework maintains comprehensive logs of user requests and access decisions, facilitating the detection of malicious activities and unauthorized access attempts. This transparency is pivotal for security monitoring and incident response.

7. **Consistent Performance Under Heavy Loads:** DLACB's processing time remains remarkably stable even when subjected to high request volumes. This analysis underscores DLACB's suitability for larger enterprises, effectively addressing intricate access control requirements.

## 4.2. DLACB Architecture

The DLACB framework uses blockchain to manage access control policies and data retrieval; it first authenticates the user and then provides the user with the appropriate level of access to the requested resources. It also keeps logs of the requests and their results, which are immutable based on the nature of the blockchain. As illustrated in Figure 2, our framework consists of four entities; user, node, decision engine, and storage.

4.1.1. **Blockchain**: A private blockchain, responsible for access control, is employed, utilizing trusted nodes and adopting the Proof-of-Authority consensus algorithm. This configuration ensures that only authorized validator nodes have the exclusive authority to validate transactions andgenerate blocks within the blockchain.

4.1.2. **User**: An entity, such as an employee, interested in storing, manipulating, or accessing data and interacts with our application.

4.1.3. **Storage**: An entity responsible for storing the user's data. The storage interacts with the blockchain when a request for storing or retrieving data is made by a user.

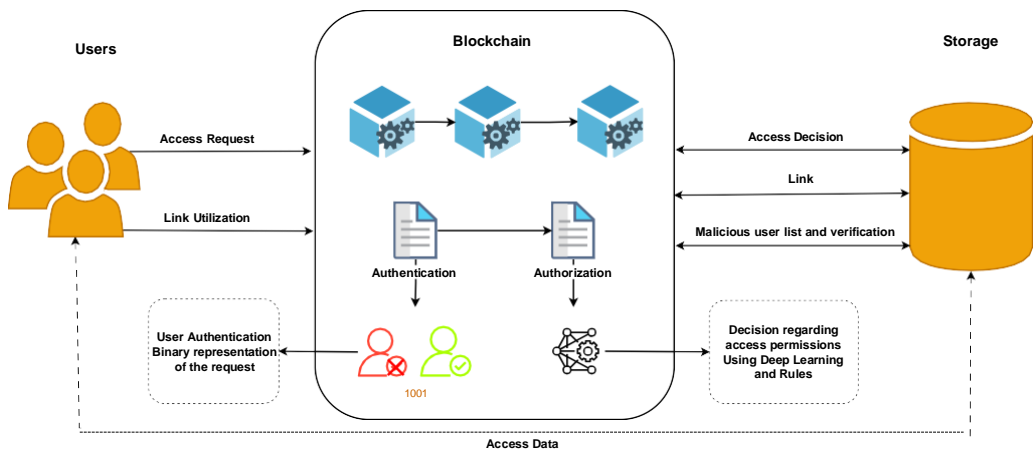

Fig. 2: DLACB Framework

## 4.3. Blockchain Structure

The DLACB framework relies on a private blockchain structure consisting of trusted validator nodes and a crucial decision engine for access control.

**Validator Node:** Validator nodes are the backbone of the DLACB blockchain. These nodes are entrusted entities that play a pivotal role in the validation of transactions and the creation of blockchain blocks. To ensure utmost trustworthiness, the network mandates that each val-idator node undergoes identity verification before participating as a trusted node. Within our framework, validator nodes are implemented as standalone, physically secured servers.

**Decision Engine:** The decision engine is the heart of DLACB's access control mechanism. It serves the critical function of user authentication and determining the access level granted to specific users, utilizing advanced deep learning models. To achieve this, the blockchain incorporates two essential smart contracts, securely stored and executed when the relevant transactions are initiated.

*Authentication Smart Contract:* This smart contract is triggered upon the reception of a data retrieval request. It rigorously validates the user's public key and confirms their net- work membership.

*Authorization Smart Contract:* Following successful authentication, the authorization smart contract takes over. This contract assesses whether the user is allowed to access a particular resource. It harnesses a deep learning model to generate an output containing a set of op- erations available for the specified resource, based on the provided inputs. Subsequently, it conducts a comparison between this output and priority rules stored within the smart contract's storage, thereby determining any specific event preferences of the organization.

Every node on the blockchain is equipped with a public and private key pair, ensuring secure transaction encryption and digital signature capabilities. To facilitate access to resources, when a user initiates a request, the blockchain undertakes the authentication and access determination process through its decision engine. Subsequently, the decision engine's result is transmitted to the storage subsystem, enabling the generation of an access link. This link is then shared with the user, granting them access to the requested resource.

## 4.4. DLACB Transactional Workflow for Data Retrieval

Within the DLACB framework, a range of transactions can be performed on the blockchain, as detailed in Table 1. These transactions serve key roles in managing access and data interactions. Consequently, the $T_{Setup}$ transaction is used to add a new node to the system, effectively expanding the blockchain's capabilities. Users employ the $T_{AccReq}$ transaction to request access to specific re- sources, specifying their data or operational requirements. On the other hand, the $T_{Link}$ transaction serves the purpose of providing users with an access link, granting them access to their requested resources. When a user accesses a resource, the $T_{Storage}$ transaction comes into play. It is utilized to log the user's interaction with the resource, ensuring the maintenance of an immutable record of data access. Finally, the $T_{Verified}$ transaction is created as an input to the smart contract within the decision engine and plays a crucial role in the authentication and authorization processes.

Table 1: Transactions

| Transaction | Parameters |
|---|---|
| $T_{Setup}$ | $pk_A, pk_U, Time, Sign_{sk_A}$ |
| $T_{AccReq}$ | $(pk_U, Time, ReqInfo), Sign_U$ |
| $T_{Link}$ | $Enc_{pk_U}(Accesslink, Nonce, Time), Sign_S$ |
| $T_{Storage}$ | $(Nonce, Time, pk_U), Sign_S$ |
| $T_{Verified}$ | $(Time, pk_{U_{binary}}, ReqInfo_{binary})$ |

The system process for utilizing the transactions for data retrieval, as depicted in Figure 3, unfolds as follows: The user initiates an *access* transaction within the blockchain, supplying their public key, request details, and freshness indicator. Subsequently, the blockchain proceeds to authenticate the user by harnessing the decision engine to determine the user's access permissions. At the core of the blockchain lies the decision engine, a fundamental component composed of two smart contracts securely stored within the blockchain. These contracts execute in response to specific transactions. When an *access* request is received, it triggers the authentication smart contract, which verifies the user's public key and confirms their network membership. Upon successful authentication, this contract generates a *verification* ($T_{Verified}$) transaction, setting in motion the authorization smart contract. The authorization smart contract evaluates whether the user is authorized to access the requested resource. More precisely, it employs a sophisticated deep-learning model explained in Section 7.2 to generate an output that enumerates permissible operations for the specified resource based on the provided inputs. This output undergoes comparison with priority rules stored within the contract's storage, determining if any specific organizational preferences exist for the event.

The result of this process is then transmitted to the storage subsystem. In response, the storage generates a $T_{Link}$ transaction containing an access link, timestamp, and nonce, all encrypted with the user's public key. This information is relayed back to the blockchain, which, in turn, forwards it to the user. Upon receiving, the users utilize the provided access link and nonce. To maintain a comprehensive record of the event, the storage subsystem generates a $T_{Storage}$ transaction within the blockchain, effectively logging the access event.

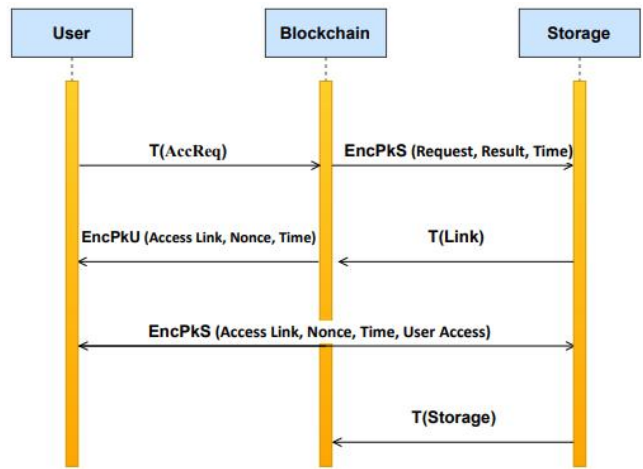

Fig. 3: Requesting Data from Storage by User

## 4.5. Malicious User Detection

At each stage of the process, the data is also sent to storage for the purpose of maintaining records. This additional step plays a crucial role in automating the identification of suspicious users within the system, thereby eliminating the need for manual blockchain queries to spot potentially harmful actions. Here's how this process operates:

1. When a user lacks proper authentication, possibly indicating their initial interaction with the system, the storage logs the user's request along with related information as suspicious activity.
2. If an authenticated user tries to access a file without the necessary permissions or attempts to exceed their authorized access limits, the storage marks this behavior as suspicious.
3. In cases where an authenticated user tries to access an incorrect file, the system records this as suspicious activity.
4. If an authenticated user repeatedly makes unsuccessful attempts to access an incorrect file, the storage system triggers an action by sending a request to the authorization smart contract on the blockchain. This request prompts the contract to adjust its predefined rules, ultimately revoking the user's access to all files within the system.

## 4.6. Enhancing Detection of Malicious Activity Queries

To improve the extraction of records related to malicious activities, two methods are employed. The first method involves recording all transactions on the blockchain, including additional information such as the user's public key and request details. This facilitates the extraction of transactions specifically associated with the targeted user or file. However, this approach may allow some activities to go unnoticed.

To address this limitation, a second method is utilized, which involves storing records in an off-chain storage system, enabling faster malicious activity detection. However, when extracting records from this storage, it is crucial to verify their integrity. To achieve this, the system utilizes the distributed Merkle root, as proposed in [5].

Whenever a request is denied by an authentication smart contract or authorization smart con-tract, the entire transaction is sent to storage along with its results and added to the list of ma- licious activities. A Merkle root is then generated and stored in the smart contract. This Merkle root is calculated by hashing the previous Merkle root (denoted as $M_{\text{prev Malicious}}$) with the trans- actions added in the current block (represented as $T_{\text{block}}$) and the output of the smart contract is Smartcontract_Output:

$$M_{\text{Malicious}} = \text{Hash}(M_{\text{prev Malicious}} \parallel (T_{\text{block}} \parallel \text{Smartcontract\_Output}))$$

By hashing the previous Merkle root with the transactions in the current block, a tamper- evident structure is created, linking the records of malicious activities. Through these two meth- ods—storing provenance records on the blockchain and using distributed Merkle roots—the sys- tem enhances its ability to extract the information of the malicious activity from the storage and verify with the blockchain.

## 5. DLACB Process and Cryptographic Techniques

In this section, we will introduce cryptographic notions and techniques, underlying protocols, the details of the transactions, and smart contracts.

## 5.1. Cryptographic Notions

We will use standard cryptographic building blocks, including asymmetric encryption, digital signatures, and a cryptographic hash function. The encryption scheme consists of a tuple (*Gen, Een, Dec*)– a generator, an encryption algorithm, and a decryption algorithm, respectively. The digital signature scheme is defined by the tuple (*Gens, Sig, V er*)– a generator, a signature, and a verification algorithm, respectively. We denote the cryptographic hash function as H and call it a hash function. Finally, we represent the blockchain memory as M; a key-value structured database that contains the complete information stored on the blockchain.

## 5.2. Smart Contract Protocols

Our approach contains an auxiliary function, *Access Verification*, and two smart contracts, *Authentication* and *Authorization*.

Algorithm 1 extracts the keys, timestamps, and signatures from the transactions. By checking the signatures and timestamps, the function establishes if the node or the user is the claimed entity and if the transaction is fresh.

---

**Algorithm 1** Access Verification Check

1: **procedure** ACCESSVERIFICATION($T_{AccReq}$, M)
2:     $a \leftarrow 0$
3:     $(pk_U, sign, time) \leftarrow Decompose(T)$
4:     **if** $H(pk_c) \in M[key]$ and $Current(time)$ and $verify(time)$ **then**
5:         $a \leftarrow 1$
6:     **end if**
7:     **return** $a$
8: **end procedure**

---

Algorithm 2 is the first smart contract performed in the decision engine. The smart contract for authentication is executed because of $T_{AccReq}$. It extracts the user key from the transaction and uses *AccessVerification()* to verify the user and determine if they were previously added and had the authority to access the system. After the user is verified, and the freshness of the request is evaluated, the smart contract converts the user data and resources into a binary representation and issues $T_{Verified}$, which triggers the authorization smart contract.

---

**Algorithm 2** Authentication

1: **procedure** VERIFICATION($T_{AccReq}$, M)
2:     $(pk_U, time, ReqInfo, sign) \leftarrow Decompose(T_{AccReq})$
3:     **if** $AccessV erificationCheck(T_{AccReq}, M)$ **then**
4:         $pk_{U_{binary}} \leftarrow binaryRepr(pk_U)$     $ReqInfo_{binary} \leftarrow binaryRepr(ReqInfo)$ 5:   $T_{Verified} \leftarrow (time, pk_{U_{binary}}, ReqInfo_{binary})$
7:         **return** $T_{Verified}$
8:     **end if**
9: **end procedure**

---

After receiving $T_{Verified}$, Algorithm 3 is executed on the decision engine to determine access according to the deep learning model for the user, and then compares the output with the priority rules and sends the final decision to the storage.

---

**Algorithm 3** Authorization

1: **procedure** DECISIONENGINE($T_{Verified}$, M)

```
2:      (time, pk_U, ReqInfo) ← Decompose(T_Verified)
3:      if Current(time) then
4:          accessList ← DecisionEngine(pk_U, ReqInfo)
5:      end if
6:      return accessList
7: end procedure
```

### 5.3. Denied Access and Malicious Activity List Verification

We have adapted the algorithm originally presented in [5] to accommodate verification alongside the storage system. The verification procedure, as detailed in Algorithm 4, is utilized to safeguard the integrity of the Malicious Activity list stored in the system. This algorithm constructs a Merkle tree and subsequently checks the result and Merkle root against the stored Merkle root. This comparison enables us to be certain whether any alterations have been made to the data.

**Algorithm 4** Enhanced Storage Verification

```
procedure STORAGEVERIFICATION(sorted malicious list)
M_Malicious ← null
for block in sorted record do
    M_block ← null
    M_block ← BlockMerkleRoot(block record)
    M_Malicious ← M_Malicious ∥ M_block
end for
stored merkle root ← Retrieve stored Merkle root
if M_Malicious matches stored merkle root then
    Data integrity verified
else
    Data integrity compromised
end if
end procedure
```

## 6. SECURITY ANALYSIS

In this section, we perform an analysis of DLACB's features that contribute to secure data sharing with users.

### 6.1. Applicable Scenarios

To evaluate the effectiveness and practicality of the proposed DLACB framework, various experiments were conducted to test its performance in different scenarios. The results of these experiments, as well as all the requests, were logged on the blockchain for record-keeping purposes. The scenarios tested were designed to cover all possible cases and include the following:

1. In the first scenario, a user who is not registered on the blockchain attempts to access a resource. In this case, the access control mechanism should deny the request, as the user's publickey is not present on the blockchain, and the system cannot authenticate the user.
2. The second scenario involved a user who is registered on the blockchain but is denied access by the deep learning model. The deep learning model in this scenario predicts that the user's request should be denied, and the system should respond accordingly.
3. The third scenario tested the case where the deep learning model approves the user's request, but the decision engine's policies reject it. This denial is based on static rules and policies defined by the organization's administrators.
4. The final scenario tested is where both the decision engine and the policies agree on authoriz-

ing the user to access the resource. This scenario represents the ideal situation where the user is authorized to access the resource.

To ensure that the testing process is comprehensive, the model was tested with all the above scenarios, and the framework was able to handle each test case successfully. For testing purposes, a dataset containing 100 users and their corresponding transactions was registered on the blockchain. However, the deep learning model recognized the entire dataset, including all users. The test cases for the scenarios were selected carefully to satisfy the description of each case and provide an accurate representation of the system's capabilities.

Moreover, the lightweight model was tested both locally and remotely, assigning a specific range of ports for users and administrators. This testing process was conducted to verify the sys-tem's functionality and performance in various settings, including both local and remote environ-ments.

## 6.2. Security Features

DLACB incorporates several robust security features to ensure the protection of sensitive data and maintain the integrity of its operations. These features work in tandem to safeguard the framework from various threats and vulnerabilities.

*Confidentiality and integrity*: DLACB's approach to ensuring confidentiality and integrity is a critical feature that helps prevent unauthorized access to sensitive data. In the DLACB blockchain, every entity has a pair of public and private keys that are used for secure communication. When requests are exchanged between two framework entities, they are encrypted with the receiver's public key, ensuring that only the intended recipient can decrypt and access the content. For example, when the storage sends a $T_{Link}$, it is encrypted with the user's public key, making it unreadable to anyone who does not possess the user's private key. Similarly, the Request Result is encrypted with the storage's public key, ensuring that only the storage can read the content. Furthermore, DLACB also ensures integrity by requiring that the entities sending a request sign the content with their private key. This signature acts as proof that the request originated from the entity and was not modified in any way. This feature helps prevent tampering with the data and ensures that all data transmitted across the network is authentic.

By combining confidentiality and integrity features, DLACB offers a robust solution for secure data sharing. The encryption of data using the recipient's public key ensures that only authorized parties can access the content, while the signature ensures that the data's integrity is maintained and it has not been tampered with.

*Preventing reply attack*: Preventing replay attacks is crucial to maintaining the security of the system. Without proper measures, an attacker could capture a message and then repeatedly send it to the receiver, causing unwanted actions to be taken. One of the primary methods used to prevent replay attacks is the use of a nonce. In DLACB, a nonce is generated for each request and produced link. A nonce is a random token that is unique for each request and is used to ensure that a malicious user cannot reuse previously communicated links. The nonce is included in the request and transaction messages, and the recipient verifies that the nonce in the message matches the nonce that they expect to receive. By including a nonce in each message, DLACB ensures that each request or transaction can only be processed once and cannot be reused by an attacker. By preventing replay attacks, DLACB maintains the integrity of the system and ensures that each request or transaction is processed only once. This makes it more difficult for an attacker to disrupt the system or gain unauthorized access to the data. The use of nonces is a common technique in blockchain systems to prevent replay attacks, and DLACB implements it effectively.

*Preventing man in the middle attack*: There are different scenarios in that a man-in-the-middle attack is possible but prevented by encryption of the data, including the nonce and the signature of the message inside the request or the transaction.

*Detecting malicious activity*: Another important feature of DLACB is its ability to detect mali- cious activity. The framework maintains a log of all user requests and the results of those requests, including which user requested the data and whether they were granted access. By logging this information on the blockchain, the system can detect any unauthorized requests or attempts to modify the data. In addition to the mentioned features, the DLACB uses PoA as a consensus algorithm, in contrast to Proof-of-Work consensus, where an attacker needs to control 51% of the network's computing power; in PoA consensus, it is necessary to control 51% of the network nodes. It is more difficult to control the nodes in a permissioned network than it is to acquire computational power, leading the attack to be more challenging in PoA [37]. PoA is scalable, efficient, and easy to implement; it can be scaled so that a blockchain using our model can have $v$ validators with $v > 2$. Validators, the trusted nodes, are the only nodes responsible for reaching consensus, while the users are free to submit transactions.

## 7. IMPLEMENTATION

This section discusses the implementation details of the introduced DLACB framework, including the blockchain and the Deep-Learning-based decision engine, and how these components work together as a system. In the end, we provide details on the experimental results and discuss the practicality of the framework from a technical perspective.

### 7.1. Blockchain Implementation

DLACB is proposed primarily for organizations that use private networks that require particular transactions, authorization procedures, and smart contracts, as discussed in section 4.2. Therefore, the blockchain is implemented from scratch, not inherited from any off-the-shelf blockchain technology, and can be considered a baseline for future frameworks using Deep Learning for the Access Control decision engine. To implement the framework as a baseline, as mentioned and as proof of concept, we use Python3 as the backend language. The framework's functionality is implemented as an API using the Flask library. In addition, a simple interface is implemented using plain HTML styled with CSS, and JavaScript is used to connect the interface and the backend by sending and receiving the HTTP requests. The implementation used for the test was an interactive web-based implementation that required the user/tester to perform each task by interacting with the web page. In the backend of this implementation, different ports are used for trusted nodes and users, and the blockchain runs on a defined port necessary for communication through the API.
The user's public and private keys are generated as pairs, and their uniqueness is ensured. Here, we use the widely used SHA-256 cryptographic hash function [45] as our encryption method for the entity and verification of digital signatures. The transactions are implemented as mentioned in section 4.2, and the transactions are then extracted and distributed among the nodes for consensus. The keys and the data are stored locally by serializing and restored by de-serializing. This method is only suitable for the demo version. However, the deployed version of the framework requires the use of a DBMS database management system such as MySQL.

### 7.2. Decision Engine Implementation

In this work, we use a machine learning model as our decision engine to allow or reject access to a resource upon a user's request. For this purpose, we use DLBAC$\alpha$ [41], a deep learning model inspired by the widely used [25] model, a convolutional Deep Neural Network successfully used in various domains. In [41], the authors used a variety of datasets, including real-world and syn-thetic datasets. The real-world datasets are the Amazon-Kaggle and Amazon-UCI datasets, which contain historical access data and access information for a large number of users and resources, respectively. Both real-world datasets are imbalanced, with the majority of tuples having grant access. The Amazon-UCI dataset is also not ABAC in nature, meaning that there are some tuples in the dataset where users with identical attribute values do not have the same access permissions. The synthetic datasets were generated with varying numbers of users, resources, user and resource metadata, and authorization tuples, each reflecting a varying degree of complexity. The synthetic datasets were designed to be more challenging than the real-world datasets, with more overlapped samples and higher complexity.

The model is implemented using popular Tensorflow and Keras machine learning libraries in Python 3.9. We use a pre-trained version of the model publicly available on GitHub. The model receives the user identification, which is an $id$, resource $id$, and returns four values for all the possible operations the user might have requested for the specific resource. The decision is made based on the requested operation. Nevertheless, the model is not limited to the use of $user_{id}$ and $resource_{id}$ and can function if trained on any attribute of the users or resources as well as user roles. The size of the trained decision engine and the weights of the model is a small 1 MB file and has little computational and memory overhead.

Deep learning models can be trained by utilizing any numerically representable feature. There- fore, Depending on the use case, the attributes utilized to train the model can be derived from attributes stored in a medical facility or, in general, an access control system containing users or resources with various attributes. Nevertheless, a deep learning model does not consider the prede- fined rules of the blockchain for accessing resources. To address this, we pass the request through a rule-checker, a set of static rules defined to ensure the validity of the decision made by the engine. The rule-checker consists of rules that apply to the user-resource pairs in specific cases.

## 8. EVALUATION

The evaluation of DLBAC$\alpha$ in [41] compared it with various access control methods, demonstrat- ing that deep learning-based techniques outperformed others. Higher F1 scores indicated improved generalization and accurate access control for unseen attributes, while higher True Positive Rate (TPR) and Precision suggested more efficient access grants. Furthermore, a lower False Posi- tive Rate (FPR) indicated better unauthorized request denials, emphasizing DLBAC$\alpha$'s superior performance in making accurate access decisions compared to other methods. The details of the discussed scores are present in the reference paper [41]. In this section, we assess the feasibil- ity of DLACB, which integrates blockchain and DLBAC$\alpha$, by evaluating the processing time of authentication and authorization smart contracts. We evaluate the processing time of both the authentication and authorization smart contracts. Additionally, we compare the processing time overhead introduced by the authorization smart contract in conjunction with the deep learning model against other access control models, RBAC and ABAC. We run all these experiments on the same computer with the specifications mentioned in the previous section. More details of this test are provided in the following subsection. To evaluate the performance in different settings, we run the smart contracts in single-thread and multi-thread modes and consider three conditions per user attributes in RBAC and six conditions per resource attributes in ABAC, for both of which these conditions are loaded from a file. For DLBAC, we assume that the weights of the model are already loaded, and only the inference time is calculated.

## 8.1. Evaluation Results

To evaluate the performance of the smart contracts mentioned before, we run many instances of these modules with different sets of inputs in different circumstances. Figure 4 provides insights into the average processing times for Authorization and Authentication smart contracts. The authorization access control corresponds to the DLACB's average time for a single request, while the Authentication access control represents the time required for a single database query.

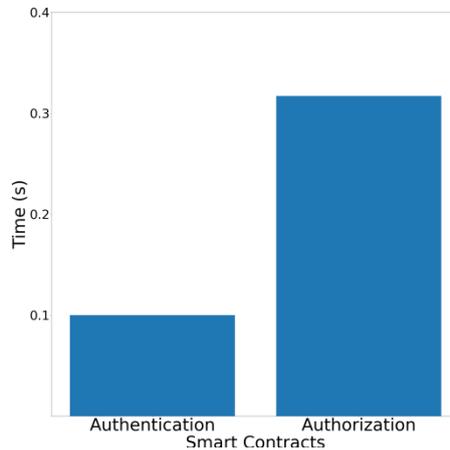

Fig. 4: Smart Contract Processing Time

In Figure 5, we conduct a comparative analysis of the access control processing times across different methodologies, encompassing deep learning-based, role-based (RBAC), and attribute-based (ABAC) access control. This figure illustrates that, unlike RBAC and ABAC, the DLBAC processing time remains relatively stable even when subjected to higher request volumes. The analysis underscores the suitability of DLBAC for larger enterprises, where the design complexity of RBAC and ABAC often leads to slower performance, especially in handling multiple concurrent requests. Conversely, in smaller organizations with fewer users and resource attributes, RBAC and ABAC might exhibit a processing time advantage.

In the multi-threaded scenario, as depicted in Figure 5, we present a graph illustrating the timetaken by different access control methods to process requests. DLBAC, the deep learning-based approach, and DLBAC' with authorization overhead are contrasted with RBAC and RBAC' for 24 and 72-core processors. Likewise, in attribute-based access control (ABAC) and ABAC', we show the processing times for 24 and 72-core processors. Notably, DLBAC demonstrates superior performance, particularly under high request loads, owing to its neural network-based nature.

In part b of the figure, which corresponds to single-threaded operation, we depict similar processing times when only a single thread is available, and requests are processed sequentially. The results are analogous, with the distinguishing feature being that DLBAC outperforms other approaches even when handling fewer concurrent requests.

## 9. DISCUSSION AND FUTURE WORK

Our study's results have demonstrated that using a pre-trained deep learning model can provide an effective solution for predicting access control policies. However, it is important to note that training the model online as the users are added to the organization is not preferred due to potential attacks on the training procedure.

To mitigate these risks, we train the model offline using the existing dataset of users and their access history and provide a pre-trained model to be used for inference. The model used here is a lightweight model and modifying it for specific applications by adding more features will not increase the required computation and storage substantially.

Looking toward the future, there are opportunities to further improve the performance of ac-cess control policy prediction models. One potential avenue for improvement is to train the model on access control patterns specific to particular industries or types of organizations. For example, models used for healthcare organizations can be designed according to the features and properties of the data available in this domain. In addition, private blockchains which are designed from scratch, can also provide advantages for organizations that require strict control over access to their data. These private blockchains can be tailored to meet the specific needs of the organization, such as implementing different consensus mechanisms or other features that are required for their particular use case. However, it is important to note that these private blockchains may not always be necessary. In cases where the needs of an organization are less specific, existing private blockchains may be sufficient for their needs. It is important to carefully evaluate the requirements of each use case and choose the appropriate solution based on those requirements.

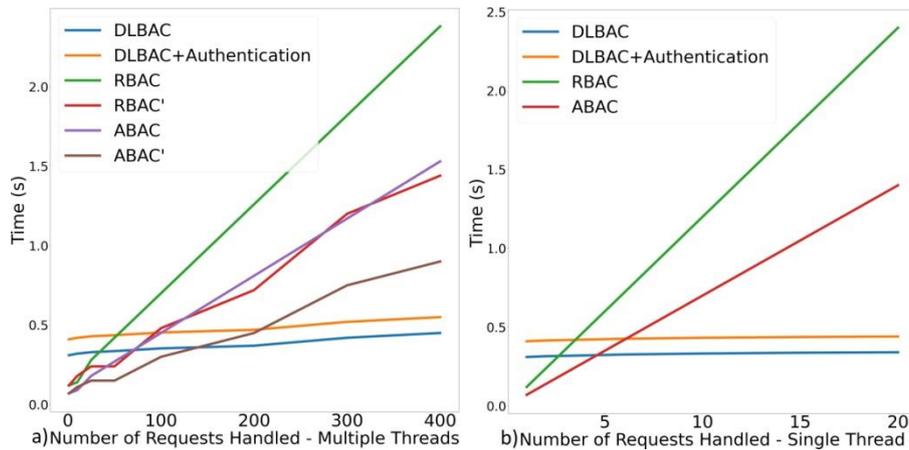

Fig. 5: Access Control Time Analysis in Authorization Smart Contract

## 10. CONCLUSION

Maintaining logs of access control patterns for detecting malicious behavior and determining the proper amount of access for the users is essential. In access control methods such as ABAC and RBAC, the administrator engineers attributes, roles, and rules for applying access control. The administrator also ensures the policies are suitable for each organization and each user has the correct amount of access to the resources, which can lead to many errors or insufficient or extensive access of some users to the resource. These methods work for predefined cases within a broader organization; therefore, they can not make the correct choice when encountering a new issue [11]. Consequently, there is a necessity for an alternative access control method.

In this paper, we presented DLACB, a blockchain-based access control platform that authenticates the users and utilizes a decision engine that employs a deep learning model along with priority rules to determine the extent of access required upon a user request. DLACB offers several unique features that contribute to its secure and efficient performance. For instance, it ensures confidentiality and integrity by encrypting requests exchanged between two framework entities with the receiver's public key. This prevents attackers from comprehending the content of a trans-action or request and ensures the authenticity of the request's origin. DLACB also prevents replay attacks by issuing a nonce, a randomly generated token, for each request and produced link, which ensures that a malicious user cannot reuse previously communicated links. Moreover, DLACB prevents man-in-the-middle attacks by encrypting the data, including the nonce and the signature of the message, inside the request or the transaction. Additionally, the platform keeps records of the user's requests to access specific data as well as their permission to access them. This process is done by logging the requests and the results on the blockchain, enabling DLACB to detect malicious requests and attacks. We also maintain a record of malicious actions within the storage system and utilize a verification algorithm to compare it against the blockchain. We deployed DLACB on a private blockchain, conducted a security analysis, and employed this framework to assess the processing time of deep learning-based access control smart contracts in comparison to traditional access control smart contract processing. The results revealed that the processing time of DLBAC remains relatively stable even when subjected to higher request volumes, further emphasizing its suitability for larger enterprises.

## ACKNOWLEDGEMENT

This work was funded by NSF grant CCF-2131509 and Augusta University Provost's office and the Translational Research Program of the Department of Medicine.

International Journal of Security, Privacy and Trust Management (IJSPTM) Vol 12, No 3/4, November 2023

## AUTHORS

**Asma Jodeiri Akbarfam** holds a Bachelor's degree from the University of Tabriz and a Master's degree in Computer Engineering from Sharif University of Technology, currently pursuing a Ph.D. in Cyber and Computer Science at Augusta University. Her research interests include blockchain, security, access control, and edge computing. 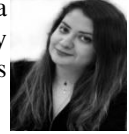

**Sina Brazandeh** received a Bachelor of Computer Engineering from Shiraz University. He is now a Master's student in Computer Engineering at Bilkent University. His research interests include Computational Biology, Bioinfor- matics, and Deep Learning. 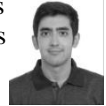

**Deepti Gupta** received her Ph.D. degree in Computer Science from the Uni- versity of Texas at San Antonio (UTSA) and also received her M.S. degree in Computer Science from UTSA. Dr. Deepti Gupta is currently an Assistant Professor at Texas A&M University-Central Texas. Dr. Gupta's research in- terests lie in the areas of security and privacy in the Internet of Things (IoT) leveraging cloud and edge computing. Her research interests also include the application of AI and Machine Learning to secure IoT and CPS infrastructures. 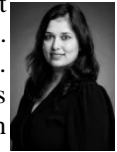

**Hoda Maleki** earned both her Bachelor's and Master's degrees at Tehran Poly- technique in Iran. In 2019, she completed her Ph.D. at the University of Con- necticut. Currently, she holds the position of Assistant Professor at Augusta University, specializing in applied cryptography and system security research. 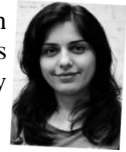